\documentclass{emulateapj}
\usepackage{epsfig}

\def\kms{\ifmmode{~{\rm km~s^{-1}}}\else{~km s$^{-1}$}\fi}
\def\lsim{\lower0.3em\hbox{$\,\buildrel <\over\sim\,$}}
\def\gsim{\lower0.3em\hbox{$\,\buildrel >\over\sim\,$}}

\def\msol{M_\odot}
\def\h2{H$_2$}
\def\flux{erg$^{-1}$ s$^{-1}$ cm$^{-2}$ Hz$^{-1}$}
\def\intensity{erg$^{-1}$ s$^{-1}$ cm$^{-2}$ Hz$^{-1}$ sr$^{-1}$}

\shorttitle{The Number of Primordial Supernovae}
\shortauthors{Wise \& Abel}

\begin{document}

\title{The Number of Supernovae from Primordial Stars in the Universe}

\author{John H. Wise and Tom Abel}
\affil{Kavli Institute for Particle Astrophysics and Cosmology,
  Stanford Linear Accelerator Center, 2575 Sand Hill Road, MS 29, Menlo
  Park, CA 94025}
\email{jwise, tabel@slac.stanford.edu}




\begin{abstract}
Recent simulations of the formation of the first luminous objects in
the universe predict isolated very massive stars to form in dark
matter halos with virial temperatures large enough to allow
significant amounts of molecular hydrogen to form. We construct a
semi-analytic model based on the Press-Schechter formalism and
calibrate the minimum halos mass that may form a primordial star with
the results from extensive adaptive mesh refinement simulations. The
model also includes star formation in objects with virial temperatures
in excess of ten thousand Kelvin.  The free parameters are tuned to
match the optical depth measurements by the WMAP satellite. The models
explicitly includes the negative feedback of the destruction of
molecular hydrogen by a soft UV background which is computed
self-consistently. We predict high redshift supernova rates as one of
the most promising tools to test the current scenario of primordial
star formation. The supernova rate from primordial stars peaks at
redshifts $\sim~20$. Using an analytic model for the luminosities of
pair-instability supernovae we predict observable magnitudes and
discuss possible observational strategies. Such supernovae would
release enough metals corresponding to a uniform enrichment to a few
hundred thousands of solar metalicity. If some of these stars produce
gamma ray bursts our rates will be directly applicable to
understanding the anticipated results from the SWIFT satellite. This
study highlights the great potential for the James Webb space
telescope in probing cosmic structure at redshifts greater than 20.
\end{abstract}

\keywords{cosmology: theory -- early universe -- supernovae: general}

\section{Introduction}
The properties of pre-reionization luminous objects are integral to
our comprehension of the process of reionization and their effect on
subsequent structure formation.  Observations of distant ($z = 6.28,
6.4$) quasars depict the relics of reionization with their
accompanying Gunn-Peterson troughs \citep{Becker02, Fan02}.
Furthermore, Lyman alpha forest carbon abundances of
$\sim$10$^{-2}$Z$_\odot$ and 10$^{-3.7}$Z$_\odot$ observed at
redshifts 3 and 5, respectively, indicate that numerous early
supernovae (SNe) enriched the intergalactic medium (IGM)
\citep{Songaila96, Songaila01}.  These early generations of stars are
at least partly responsible for ionizing the Universe.  A fraction of
these stars lie in protogalaxies, but the other fraction of early
stars are metal-free, form through molecular hydrogen cooling and are
very massive (M $\sim$ 100$\msol$) \citep{Abel00, Abel02}.  In
$\Lambda$CDM cosmologies, these stars form at 10 $\lsim$ z $\lsim$ 50.
The Wilkinson Microwave Anisotropy Probe (WMAP) data further
constrains the epoch of reionization from the measurement of the
optical depth due to electron scattering, $\tau_{es}$ = 0.17 $\pm$
0.04 (68\%), which corresponds to a reionization redshift of 17 $\pm$
5 when assuming instantaneous reionization \citep{Kogut03}.  However
in this paper, we shall show that this epoch is gradual and its effect
on primordial star formation.

In hierarchical models of structure formation, small objects merge to
form more massive structures.  Eventually, a fraction of halos are
massive ($\sim 5 \times 10^5\msol$ at $z \sim 20$) enough to host
cooling gas \citep{Couchman86, Tegmark97, Abel98, Fuller01}.  These
halos do not cool through atomic line cooling since $T_{vir} < 10^4$;
however, primordial gas contains a trace of \h2.  Free electrons which
allow molecular hydrogen to form that acts as an effective coolant at
several hundred degrees through rotational and vibrational
transitions.  \h2 is easily photo-dissociated by photons in the
Lyman-Werner (LW) band, 11.26--13.6eV \citep{Field66, Stecher67}, thus
\h2 can be destroyed by distant sources in a neutral Universe.
Primordial stars produce copious amounts of UV photons in the LW band,
destroying the most effective cooling process at $z \sim 20$.  This
negative feedback from the UV background significantly inhibits the
primordial star formation rate in the early universe by requiring a
larger potential well for gas to condense.  Many groups have explored
the effects of a UV background on cooling and collapsing gas.
Firstly, \citet{Dekel87} discovered that \h2 can be dissociated from
large distances.  Then more quantatively, \citet{Haiman97b} questioned
how the collapse homogeneous, spherical clouds are affected by a UV
background.  In more detail, \citet{Haiman00} determined whether a gas
cloud collapsed by comparing the cooling time with the current
lifetime of the cloud, which was calculated in the presence of solving
the spherically symmetric radiative transfer equation along with
time-dependent \h2 cooling functions.  However, these studies only
considered spherically symmetric cases while realistically these halos
are overdensities within filaments.  To combat this problem,
\citet{Machacek01} employed a three-dimensional Eulerian adaptive mesh
refinement (AMR) simulation to determine quantitative effects of a UV
background on gas condensation.

We consider lower mass stars to form in more massive halos that may
fragment via atomic line cooling as well as the metal lines from the
heavy elements expelled by the earlier generation of Pop III stars.
We use the prescription outlined in \citet{Haiman97a} (hereafter HL97)
to model the abundances and luminosities of these stars.

With infrared space observatories, such as the Spitzer Space
Telescope, Primordial Explorer \citep[PRIME;][]{Zheng03}, and James
Webb Space Telescope (JWST), sufficient sensitivity is available to
detect the SNe from primordial stars.  Although SNe remnants are
bright for short periods of time, they may be the best chance to
directly observe primordial stars due to their large intrinsic
luminosities.  If these events are recorded, many properties, such as
mass, luminosity, metallicity, and redshift, of the progenitors can be
calculated using metal-free or ultra metal-poor SNe models, which can
validate or falsify simulations of the first stars \citep{Abel02}.  To
evolve our calculation through redshift, we need to retain information
about the UV background and number densities of primordial stars,
which can extend to determining such quantities as volume-averaged
metallicity and ionized fraction of the Universe.  Our model is
constrained with (a) the WMAP optical depth measurement, (b)
primordial star formation and suppression in high resolution
hydrodynamical simulations, and (c) local observations of dwarf
galaxies that constrain high redshift protogalaxies properties.

We organize the paper as follows.  In \S2, we report the
semi-empirical method behind our calculations, which incorporates
effects from negative feedback and an ionized fraction and is
constrained from the WMAP result and local dwarf galaxy observations.
In \S3, we present the number density of primordial SNe in the sky and
the evolution of SNe rates, metallicity, and related SNe quantities.
We also explore the feasibility of observing these SNe by calculating
their magnitudes and comparing them with the sensitivities of infrared
space observatories.  Then in \S4, we suggest possible future
observations and numerical simulations that could further constrain
our model.  Finally, \S5 discusses and summarizes our results and the
implications of the first observations of SNe from primordial stars.

\section{The Method}
We use several theories and results from simulations of structure
formation and metal-free stars.  We use a $\Lambda$CDM cosmology with
$\Omega_\Lambda$ = 0.70, $\Omega_{\rm{CDM}}$ = 0.26, $\Omega_b$ =
0.04, h = 0.7, $\sigma_8$ = 0.9, and n = 1.  $\Omega_\Lambda$,
$\Omega_{\rm{CDM}}$, and $\Omega_b$ are the fractions of mass-energy
contained in vacuum energy, cold dark matter, and baryons,
respectively.  $h$ is the Hubble parameter in units of 100 km s$^{-1}$
Mpc$^{-1}$.  n = 1 indicates that we used a scale-free power spectrum,
and $\sigma_8$ is the variance of random mass fluctuations in a sphere
of radius 8h$^{-1}$ Mpc.  We use a CDM power spectrum defined in
\citet{Bardeen86}, which depends on $\sigma_8$ and h.

The remaining parameters are the primordial stellar mass, and the
factors f$_{esc}$ and f$_\star$, which dictates the production and
escape of ionizing photons, where f$_{esc}$ is the photon escape
fraction, and f$_\star$ is the star formation efficiency.

To evolve the ionization and star formation behaviors of the Universe,
we must determine the density of dark matter halos that host early
stars.  In $T_{vir} < 10^4$ K halos (henceforth ``minihalos''), \h2
cooling is the primary mechanism that provides means of condensation
into cold, dense objects.  However in $T_{vir} > 10^4$ K, which
corresponds to $M_{vir} > 10^8 \msol [(1+z)/10]^{-3/2}$, halos,
hydrogen atomic line cooling allows the baryons to fragment and cool
into stars.

\subsection{Minihalo Star Formation}

The first quantity we need to begin our calculation is the minimum
halo mass that forms a cold, dense gas core due to \h2 cooling.
Radiation in the LW band photo-dissociates \h2, which inhibits star
formation in minihalos.  This negative feedback from a UV background
does not necessarily prohibit the formation of primordial stars in
minihalos, but only increases the critical halo mass in which
condensation occurs, and delays the star formation.  From their
simulations of pre-galactic structure formation, \citet{Machacek01}
determined the minimum mass of halo that hosts a massive primordial
star is
\begin{equation}
\label{minMass}
\frac{M_{min}}{\msol} = \exp\left(\frac{f_{cd}}{0.06}\right)
  \left(1.25 \times 10^5 + 8.7 \times 10^5
  F_{LW,-21}^{0.47}\right),
\end{equation} 
where $M_{min}$ is the minimum halo mass that contains a cold, dense
gas core, $f_{cd}$ is the fraction of gas that is cold and dense, and
$F_{LW}$ is the flux within the LW band in units of 10$^{-21}$ \flux .
For our calculation, we consider $f_{cd} = 0.02$, which is a
conservative estimate in which we have an adequate source of star
forming gas.  With our chosen $f_{cd}$ and no UV background, a 1.74
$\times$ 10$^5$ $\msol$ halo will form a cold, dense core, which will
continue to form a primordial star.  In a typical UV background of J =
10$^{-21}$ \intensity , the minimum mass is 4.16 $\times$ 10$^6
\msol$.  Additionally, minihalos can only form a star within neutral
regions of the Universe since they are easily photoionized
\citep{Haiman01, Oh03} and \ion{H}{1} is a necessary ingredient for
producing \h2.

Numerical simulations \citep[e.g.][]{Abel02, Bromm02a} illustrated
that fragmentation within the inner molecular cloud does not occur and
a single massive (M $\sim$ 100$\msol$) star forms in the central
regions.  These stars produce hard spectra and tremendous amounts of
ionizing photons.  For example, the ionizing photon to stellar baryon
ratio n$_\gamma$ $\sim$ 91300, 56700, and 5173 for H, He, He$^+$ in a
200$\msol$ star \citep{Schaerer02}.  We calculate the ionizing photon
flux by considering
\begin{equation}
\label{ion_mini}
\left(\frac{dN_\gamma}{dt}\right)_{mini} = \bar{Q} \: T_{life} \:
\frac{d\rho_{mini}}{dz} \: \frac{dz}{dt},
\end{equation}
where Q is the time-averaged photon flux, T$_{life}$ is the stellar
lifetime, and $\rho_{mini}$ is the comoving number density of
minihalos that we determine from an elliposdial variant of
Press-Schechter (PS) formalism \citep{PS74, Sheth02}.

The primordial initial mass function (IMF) is unknown, therefore, we
assume a fixed primordial stellar mass for each calculation.  We run
the model for primordial stellar masses, M$_{fs}$, of 100, 200, and
500 $\msol$ and use the time-averaged emissivities from metal-free
stellar models with no mass loss \citep{Schaerer02}.  In our models,
the minihalo is quickly ionized and all photons escape into the IGM
\citep{Whalen04}.  Furthermore, we use a blackbody spectrum at 10$^5$
K to approximate the spectrum of the primordial star to consider the
flux in the LW band since surface temperatures are virtually
independent of mass at M $\gsim$ 80$\msol$.

\subsection{Star Formation in Protogalaxies}
In $T_{vir} > 10^4$ K halos, baryons can cool efficiently through
atomic line cooling, thus fragmenting and forming stars.  As described
in HL97, we parameterize the properties of these stars by a couple of
factors.

\subsubsection{Star formation efficiency}
High redshift galaxies appear to have similar properties as local
dwarf galaxies.  We can constrain the star formation efficiency
f$_\star$ by letting local observations guide us.  In these galaxies,
f$_\star$ range from 0.02 to 0.08 \citep{Taylor99, Walter01}.  Using
the orthodox Schmidt star formation law, \citet{Gnedin00} estimated
f$_\star$ to be 0.04 if it was constant over the first 3 Gyr but can
also be as low as 0.022.  It should be noted that f$_\star$ can be
higher if star formation ceased after a shorter initial burst.

Analyses of metallicities in local dwarf galaxies reveal their prior
star formation.  Both Type Ia and Type II SNe contribute iron to the
IGM, but Type II SNe provide most of the $\alpha$-process elements
(e.g. C, N, O, Mg) to the IGM.  Type II SNe occur on timescales $< 3
\times 10^7$ yr while Type Ia are delayed by $3 \times 10^7$ yr to a
Hubble time \citep{Matteucci01}.  Therefore, we expect an
overabundance of $\alpha$-process elements with respect to iron.
[$\alpha$/Fe] versus [Fe/H]\footnote{We use the conventional notation,
[X/H] $\equiv$ log(X/H) - log(X$_\odot$/H$_\odot$)} plots help us
inspect the evolution of the ISM/IGM metallicity.  If the prior star
formation is inefficient (spirals, irregulars), [$\alpha$/Fe] is only
shortly overabundant, which is characterized by a short plateau versus
[Fe/H].  On the other hand, if the star formation is fast and occurs
early in the lifetime of the galaxy, [$\alpha$/Fe] remains in the
plateau longer due to the quick production of metals within Type II
SNe \citep[see Figure 1 in][]{Matteucci02}.  Recently, \citet{Venn03}
discovered no apparent plateau in [$\alpha$/Fe] versus [Fe/H] in dwarf
spheroidal and irregular galaxies.  They conclude that star formation
must have been on timescales longer than Type Ia SNe enrichment, which
hints at a low and continuous star formation rates in these
protogalaxies when compared to recent star formation.

These low star formation efficiencies are further supported by the
galaxies contained in the Sloan Digital Sky Survey
\citep[SDSS;][]{York00}.  Star formation efficiencies within low mass
galaxies (M $<$ 3 $\times$ 10$^{10} \msol$) decline as M$^{2/3}$
\citep{Kauffmann03}.  At high redshift and before reionization, most
of the protogalaxies tend to be few $\times$ 10$^7$ $\msol$, which is
comparable to many local dwarf galaxies \citep{Mateo98}.  It should
also be noted that even within individual Local Group galaxies no star
formation history is alike, but it is worthwhile to adopt a global
star formation efficiency and observe the consequences on reionization
and primordial star formation.  With the stated constraints, we take
f$_\star$ = 0.04 in our main model in concordance with the Schmidt Law
\citep{Gnedin00} and stellar abundances \citep{Venn03}.


\begin{deluxetable}{ccccc}
\label{imf}
\tablecolumns{5}
\tabletypesize{}
\tablewidth{0pc}
\tablecaption{Ionizing photon production per stellar baryon and
luminosities for metal-poor IMFs}
\tablehead{\colhead{Z} & \colhead{n$_{\gamma,\:H}$} &
\colhead{n$_{\gamma,\:He}$} & \colhead{n$_{\gamma,\:He^+}$} &
\colhead{log $\cal{L}$}\\
\colhead{} & \colhead{} &\colhead{} &\colhead{} & \colhead{[erg s$^{-1}$ M$_\odot^{-1}$]}}
\startdata
10$^{-7}$	& 20102	& 8504	& 8	& 36.20\\
10$^{-5}$	& 15670	& 5768	& 0.2	& 36.16\\
0.0004	        & 13369	& 3900	& 0	& 36.28\tablenotemark{a}
\enddata
\tablenotetext{a}{See text for a discussion on values.}
\end{deluxetable}

\subsubsection{Stellar luminosities}
We considered several IMFs for star formation within protogalaxies.
Our main model uses a Salpeter IMF with a slope $\alpha$ = --2.35,
metallicity Z = 10$^{-7}$, 10$^{-5}$, and 0.0004, and (M$_{low}$,
M$_{up}$) = (1, 100)$\msol$.  We take the continuous starburst
spectrum evolution from this particular IMF that was calculated with
Starburst99 \citep{Schaerer03, Leitherer99}.  To estimate the ionizing
photon per stellar baryon ratio, n$_\gamma$, and luminosities of the
IMF at a particular metallicity, we interpolate in log$_{10}$ space
between the two adjacent IMFs.  When Z $<$ 10$^{-7}$, then we consider
the Z = 10$^{-7}$ IMF.  The properties of these IMFs are listed in
Table 1.  Note that more metal-rich starbursts result in a softer
spectrum and less ionizing photons in which n$_{\gamma,\:H}$ decreases
almost by a factor of 2.

In the Z = 0.0004 IMF, we retain the luminosity from the Z = 10$^{-5}$
and set n$_{\gamma,\:He^+}$ = 0.  We choose to do so because of the
theoretical uncertainty of hot Wolf-Rayet (WR) stars and their
presence \citep[for a review, see][]{Schaerer00}.  In
\citet{Schaerer03}, the luminosity and spectral hardness increases due
to the presence of hot Wolf-Rayet (WR) stars at higher metallicities.
However, \citet{Smith02} calculate hot WR spectra that are
significantly softer.

\subsubsection{Ionizing photon escape fraction}
Radiation emitted by these stars have a probability $f_{esc}$ to
escape from the protogalaxy and ionize the IGM.  The protogalaxy ISM
density and composition plays the biggest role in determining this
factor.  \citet{Heckman01} showed that local and distant starburst
galaxies, including the gravitationally lensed galaxy MS 1512-cB58 (z
= 2.7), have $f_{esc}$ $\lsim$ 0.06.  This result agrees with previous
analyses of the $f_{esc}$ in the Lyman continuum \citep{Leitherer95,
Hurwitz97}.  However, Lyman break galaxies may have higher $f_{esc}
\gsim 0.2$ \citep{Steidel01}.  In conjunction with f$_\star$ = 0.04
and M$_{fs}$ = [100, 200, 500]$\msol$, the choice of f$_{esc}$ =
[0.050, 0.033, 0.028], respectively, in our calculation results in the
same optical depth from WMAP, $\tau_{es}$ = 0.17, thus we use these
values in the main models.  \citet{Wood00} and \citet{Ricotti00} have
argued that the escape fraction may be very small due to the much
higher densities at high redshift. However, the shallow potential
wells and their small size make them also susceptible to
photo-evaporation and effects of radiation pressure
\citep{Haehnelt95}.  Other uncertainities such as dust content,
metallicity, and whether the ISM density scales as (1+z)$^3$ blur our
intuition about the escape of ionizing radiation.  Therefore, we allow
$f_{esc}$ to vary from 0.001--0.25 since it is unclear whether
high-redshift, low-mass protogalaxies allow photons to escape due to
self-photoevaporation or absorb the photons due to a higher proper gas
density when compared to starburst galaxies.

\subsubsection{Ionizing Photon Rates}
These factors are multiplicative in amount of radiation that is
available from protogalaxies to ionize the IGM.  The rate of
photons emitted that can ionize species X is 
\begin{equation}
\label{dn_dt}
\left(\frac{dN_\gamma}{dt}\right)_{proto} = \rho_0 (1+z) \frac{f_{esc}
\: f_\star \: n_{\gamma,X}}{\mu \: m_p} \: \frac{d\psi_{proto}}{dz}
\frac{dz}{dt},
\end{equation}
where $\mu$ is the mean molecular weight, $\rho_0$ =
$\Omega_b$(3H$_0^2$/8$\pi$G), $m_p$ is the mass of a proton, and
$\psi_{proto}$ is the mass fraction contained in protogalaxies that is
calculated by PS formalism.  We restrict the product of $f_{esc}$ and
$f_\star$ to be in a range from 10$^{-4}$--10$^{-2}$ since the
resulting reionization histories fall within the measured $\tau_{es}$.
The reionization epoch greatly depends on the factors $f_{esc}$ and
$f_\star$.  In order to explore the consequences of different values
of $\tau_{es}$ and their resulting SNe rates, we vary the factors
f$_{esc}$ and f$_\star$.
\begin{figure}[!b]
\vspace{0.25cm}
\plotone{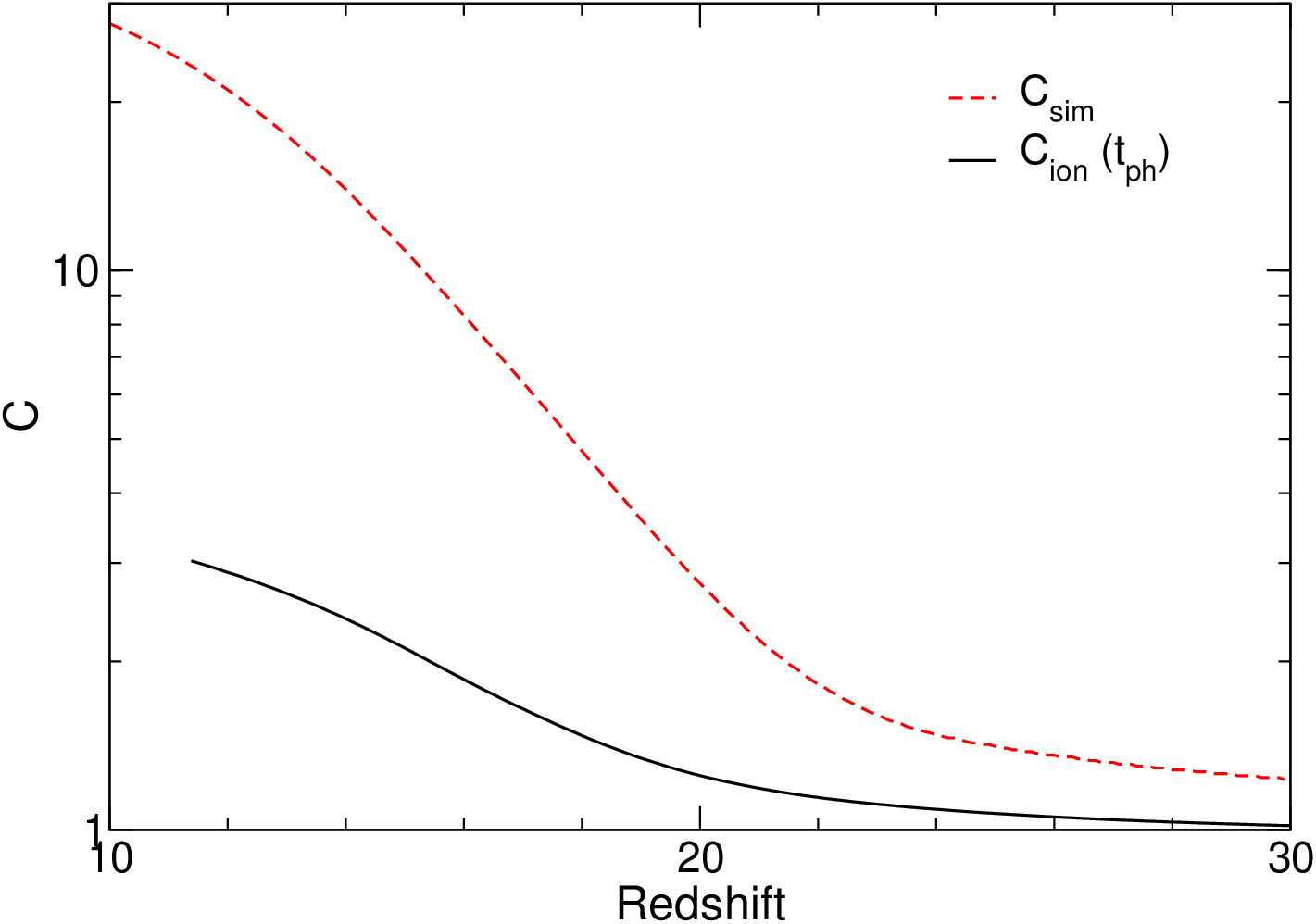}
\caption{\label{clumping}
The dashed line is the gas clumping factor as calculated in
our adiabatic hydrodynamical AMR simulation.  Using equation
(\ref{clumpchange}), we estimate the clumping factor in the ionized
region to be solid line.}
\end{figure}
\begin{figure*}[!t]
\resizebox{\textwidth}{!}{\rotatebox{0}{\plotone{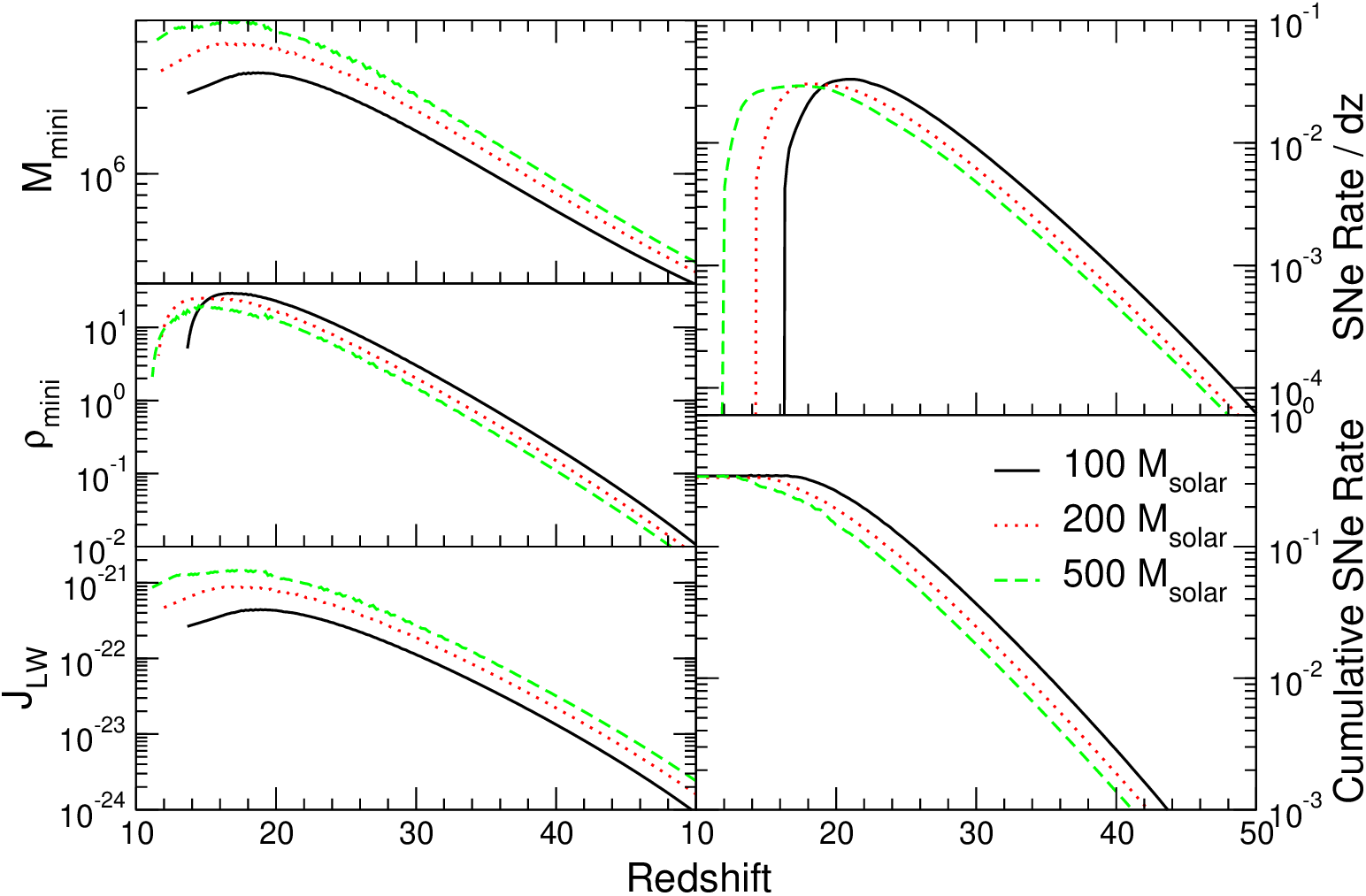}}}
\caption{\label{evo_plot} ({\em Clockwise from upper right})
Primordial SNe rates (yr$^{-1}$ deg$^{-2}$) per unit redshift;
Cumulative primordial SNe rate (yr$^{-1}$ deg$^{-2}$); Specific
intensity (\intensity) in the LW band; Comoving density (Mpc$^{-3}$)
of halos above the critical star formation mass in neutral regions;
Critical halo mass ($\msol$) for primordial star formation.  The
solid, dotted, and dashed lines correspond to calculations run with a
fixed primordial stellar mass of 100, 200, and 500 $\msol$,
respectively.}
\end{figure*}
\subsection{Clumping Factor}
Overdense regions experience an increased recombination rate.
Overdensities are characterized by the gas clumping factor C =
$\langle n_H^2 \rangle / \langle n_H \rangle^2$, and the recombination
rate is increased by this factor.  To calculate this parameter in our
model, we utilize the same cosmological Eulerian AMR code {\em enzo}
\citep{Bryan99} as in \citet{Machacek01} in a 256$^3$ simulation with
a comoving box side of 500kpc, eight levels of refinement, and the
previously specified cosmological parameters.  A total (gas) mass
resolution of 1013 (135) $\msol$ results from the resolution in the
top grid, which is smaller than the cosmological Jeans mass
(\ref{jeans}).  Thus, we account for all collapsed halos in our
clumping factor calculation.  The simulation is purely adiabatic with
no background radiation or atomic/molecular cooling.  We show
C$_{sim}$ in Figure \ref{clumping}.  Cooling only affects localized
regions of star formation and does not contribute greatly in the
boosting of C.  However, ionizing radiation causes photoevaporation of
halos, which decreases C in the process \citep{Haiman01}.  We correct
for photoevaporation by considering all minihalos with M$_J$ $<$ M $<$
M$_{min}$ are photoevaporated in the ionized regions of the Universe,
where
\begin{equation}
\label{jeans}
M_J \approx 10^4 \left(\frac{\Omega_M h^2}{0.15}\right)^{-1/2} 
	\left(\frac{\Omega_b h^2}{0.02}\right)^{-3/5}
	\left(\frac{1+z}{11}\right)^{3/2} \msol
\end{equation}
is the cosmological Jeans mass.  For simplicity, we assume the IGM has
a clumping factor of unity although underdense regions in the IGM
correspond to C$_{IGM}$ $\lsim$ 1.  We concentrate on the gas clumping
factor in the ionizing regions since the goal is to calculate the
increase in recombination rates.  Consider an ionized region whose
volume filling fraction F$_H$ is increasing.  The expansion will
incorporate unaffected, clumpy material into the region.  The first
term in (\ref{clumpchange}) accounts for this effect.  Any
overdensities in the ionized region will be photoevaporated in
approximately a sound crossing time of the halo,
\begin{equation}
t_{ph} \approx \beta\frac{R_{vir}}{10 \kms},
\end{equation}
where $\beta$ is a normalization factor that accounts for the
differences in halo densities at various redshifts and masses and
R$_{vir}$ is the mean virial radius of halo with M$_J$ $<$ M $<$
M$_{min}$ \citep{Haiman01}.  Since $\beta$ remains within $\sim$15\%
of unity for minihalos, we infer $\beta$ = 1.  The photoevaporation of
overdensities is the second term in the evolution of the gas clumping
factor, which is
\begin{equation}
\label{clumpchange}
\dot{C} = C_{sim}\dot{F}_H - \frac{C-1}{t_{ph}}.
\end{equation}
The effect from photoevaporation is illustrated in Figure
\ref{clumping}.  Overdensities that are engulfed by ionized regions
cannot sufficiently increase the gas clumping factor to overcome the
photoevaporation that occurs, and C remains within the range 1--3 for
the entire calculation in our main model.  This equation is weakly
sensitive to t$_{ph}$ since varying t$_{ph}$ by a factor of 6 alters C
only by a factor of 2.  In a later paper, we shall computationally
address the effect of radiation on the clumping factor.

\subsection{The Evolution of UV Background and Halo Densities}
We initialize the following method at z = 75 with no UV background,
evolve the cosmological radiative transfer equation (\ref{cosmo_rte})
at the questioned redshift, and repeat the described procedure until
cosmological hydrogen reionization occurs.

Given a minimum mass of a star-forming halo, we can exploit PS
formalism to calculate number densities of these halos.  For
minihalos, we calculate their number densities for halos with masses
above M$_{min}$ (\ref{minMass}) and virial temperatures below 10$^4$
K.  Likewise, the protogalaxy mass fraction is calculated by
considering all halos more massive than a corresponding T$_{vir}$ =
10$^4$ K.

We choose a variant of PS formalism, which is an ellipsoidal collapse
model that is fit to numerical simulations \citep{Sheth99, Sheth01,
Sheth02}.  This model is concisely summarized in \citet{Mo02}.  In
minihalos, it is reasonable to assume one star forms per halo since
E$_{bind}$ $\lsim$ E$_{SNe}$, where E$_{bind}$ and E$_{SNe}$ are the
binding energy of the host halo and kinetic energy of the primordial
SNe, respectively.  The gas is totally disrupted in the halo and
requires $\sim$100Myr $\ll$ t$_H$, where t$_H$ is the Hubble time, to
re-collapse (Abel, Bryan, \& Norman 2002, unpublished).  On main
sequence, primordial stars evacuate large Str\"{o}mgren spheres on the
order of kpcs \citep{Whalen04}.  Therefore,
\begin{equation}
\label{rho_star}
\frac{d\rho_\star}{dz} = \frac{d\rho_{mini}}{dt} \frac{dt}{dz},
\end{equation}
where $\rho_\star$ is the comoving density of primordial stars.  With
that in mind, we can calculate the volume-averaged emissivity
(\ref{ems_eqn3}) of the Universe by using metal-free stellar models
\citep{Schaerer02} and the prescription outlined in \S2.2 for
minihalos and protogalaxies, respectively.

We evolve the spectrum from early stars to investigate how the UV
background behaves, particularly in the LW band, with the cosmological
radiative transfer equation,
\begin{equation}
\label{cosmo_rte}
\left( \frac{\partial}{\partial t} - \nu H \frac{\partial}{\partial
  \nu} \right) J = -3HJ - c\kappa J +
  \frac{c}{4\pi}\epsilon,
\end{equation}
where $H = H(z) = 100h E(z)$ is the Hubble parameter, and $E(z) =
\sqrt{\Omega_\Lambda + \Omega_m (1+z)^3}$.  $J = J(\nu,z)$ is specific
intensity in units of \intensity.  $\kappa$ is the continuum
absorption coefficient per unit length, and $\epsilon = \epsilon(\nu)$
is the proper volume-averaged emissivity \citep{Peebles93}.  The
emissivity will be the current total luminosity per unit volume of
early stars, which can be expressed as
\begin{equation}
\label{ems_eqn1}
\epsilon_{mini}(\nu) = \rho_{mini} \times L(\nu) \times f(\nu)
   \times \left(1 - \frac{n_{HII}}{n_H}\right),
\end{equation}
\begin{equation}
\label{ems_eqn2}
\epsilon_{proto}(\nu) = \nu^{-1} \rho_0 \frac{d\psi_{proto}}{dt}
\times L(\nu, f_\star) \times f(\nu),
\end{equation}
\begin{equation}
\label{ems_eqn3}
\epsilon(\nu) = (1+z)^3 \left[\epsilon_{mini}(\nu) +
\epsilon_{proto}(\nu)\right],
\end{equation}
where f (\ref{lyman_abs}) is a factor that accounts for absorption
from Lyman series transitions.  L is the luminosity of the object and
is defined as
\begin{displaymath}
L(\nu) = \left\{ \begin{array}{r@{\quad}l} 
4\pi R^2 B_\nu(T=10^5K) & {\rm (mini)}\\
f_{O\star} f_\star \mathcal{L} \frac{B_\nu(T=23000K)}{B_\nu(\nu = 2.7kT/h_p; \: T=23000K)} & {\rm (proto)}
\end{array} \right.,
\end{displaymath}
where B$_\nu$ is a blackbody spectrum, and $\cal{L}$ is listed in
Table 1.  R is the radius of the primordial star \citep{Schaerer02}.
f$_{O\star}$ is the fraction of O stars in the starburst
\citep{Schaerer03}.  k and h$_{\rm{p}}$ are Boltzmann's constant and
Planck's constant, respectively.  For the protogalaxies, T $\sim$
23000K because the spectrum in the LW band will be dominated by OB
stars, and we weight the luminosity by this blackbody spectrum.  This
effect and the increasing UV background effectively squelches minihalo
star formation.  Emissivity will be nearly zero above 13.6eV in the
neutral Universe due to the hydrogen and helium absorption.

Photons from primordial stars and protogalaxies between 13.6eV and
several keV ionize the surrounding neutral medium.  Using the
intrinsic $\geq$13.6eV ionizing photon rates from primordial stars
(\ref{ion_mini}) and protogalaxies (\ref{dn_dt}), we calculate a
volume-averaged neutral fraction.  The change of ionized hydrogen
comoving density due to photo-dissociation and recombination is
\begin{figure}[!t]
\vspace{0.75cm}
\plotone{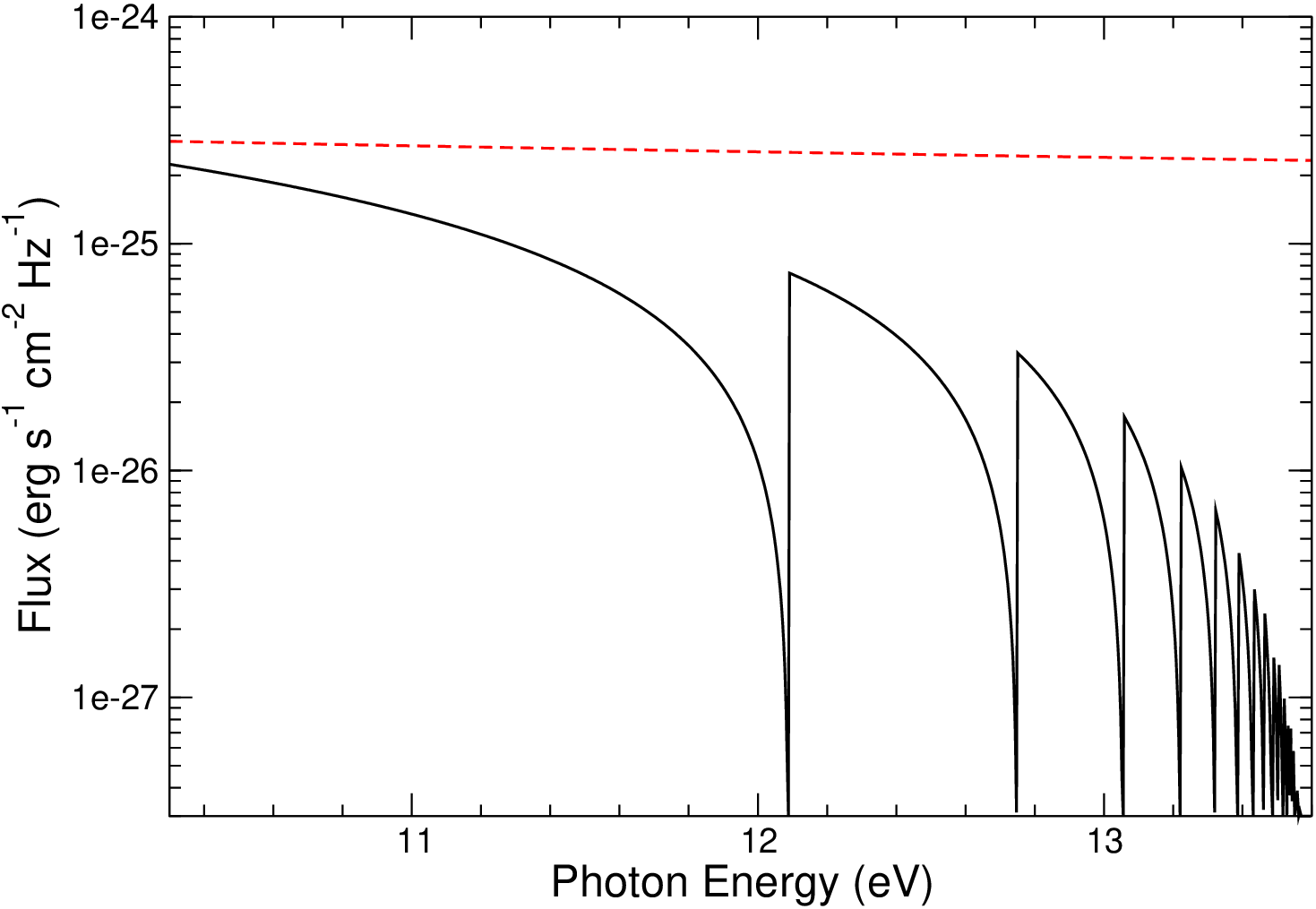}
\caption{\label{saw}
The red dashed line is an example unprocessed spectrum of a
continuous source of radiation.  After being absorbed and re-emitted
by Lyman-series transitions, it transforms into the ``sawtooth''
spectrum ({\em black solid line}).  This particular spectrum is for
z$_{on}$ = 30 and z$_{obs}$ = 20.}
\end{figure}
\begin{equation}
\label{neutral_h}
\frac{dn_{HII}}{dt} = -\frac{dN_\gamma}{dt} + C k_{rec} (1+z)^3 n_H^2
\left(1-\frac{n_{HII}}{n_H}\right)^2,
\end{equation}
where n$_{HII}$ and n$_H$ = 0.76$\rho_0$/m$_p$ are the ionized and
total hydrogen comoving density, C is the clumping factor, and
k$_{rec} =$ 2.6 $\times$ 10$^{-13}$ s$^{-1}$ cm$^{-3}$ is the case B
recombination rate of hydrogen at T $\approx$ 10$^4$.

In the case for helium, $k_{rec}$ is approximately equal to the
hydrogen case, but since the helium number density is less than
hydrogen, less recombinations occur.  Na\"{\i}vely, this will result
in a higher He$^+$ fraction than H$^+$ due to the hardness of the
primordial radiation.  Realisticly when the ionizing photons reach the
ionization front, they will ionize either helium or hydrogen since
hydrogen still has a finite photo-ionization cross-section above
23.6eV.  We consider the He$^+$ regions to be equal to the H$^+$
regions.  Then we add N$_{\gamma,\:He}$ to N$_{\gamma,\:H}$ to
compensate for this effect.  We perform a similar analysis on the
double ionization of helium in which k$_{rec,\: He^+}$ = 1.5 $\times$
10$^{-12}$ s$^{-1}$ cm$^{-3}$ but we evaluate equation
(\ref{neutral_h}) directly with N$_{\gamma, \:He^+}$.
\begin{figure}[!t]
\begin{center}
\vspace{0.75cm}
\plotone{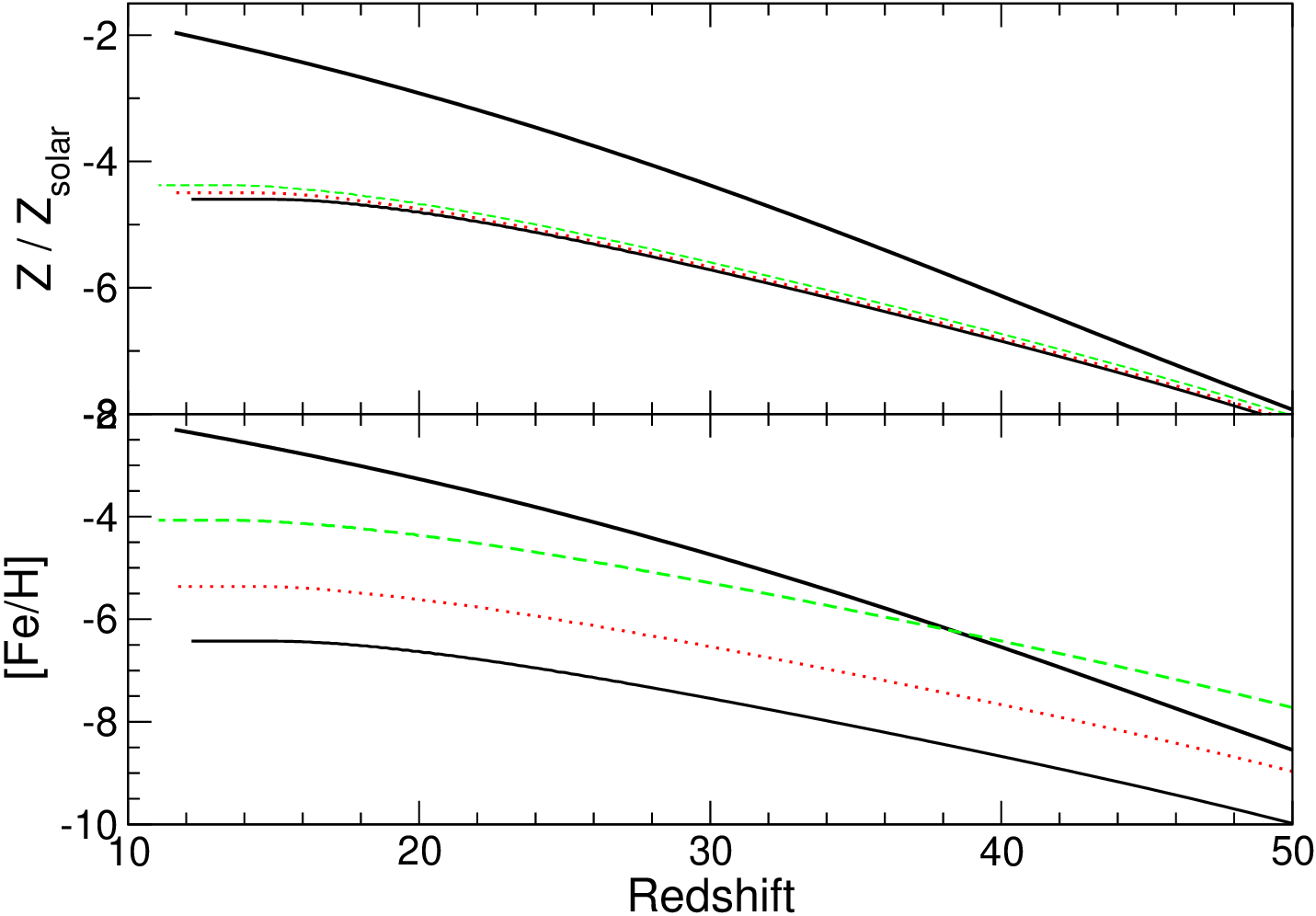}
\caption{\label{metallicity}
({\em Top graph}) The thick solid line is the volume-averaged
metallicity from protogalaxies and primordial stars in the 200$\msol$
model.  The thin dotted, solid, and dashed lines are volume-averaged
metallicities from 175, 200, 250 $\msol$ primordial stars only.  ({\em
Bottom graph}) Iron abundance relative to solar.  Legend is the same
as the top graph.  The large spread in iron from primordial stars
arises from the steep dependence of iron ejecta versus mass in pair
instability SNe.}
\end{center}
\end{figure}

The absorption coefficient in Equation (\ref{cosmo_rte}) is ignored
since we take into account the Lyman series line absorption by the
following procedure.  Photons with 11.26eV $<$ E $<$ 13.6eV escape
into the IGM, which will photo-dissociate \h2.  To calculate the flux
within the LW band, we must consider the processing of photons by the
Lyman series transitions \citep{Haiman97b}.  Before reionization,
these transitions absorb all photons at their respective energies.
These absorbers can be visualized as optically thick ``screens'' in
redshift space, for which the photon must have been emitted after the
farther wall in redshift.  This process creates a sawtooth spectrum
with minimums at redshift screens, which is illustrated in Figure
\ref{saw}.  For instance, an observer at $z = 20$ observes photons at
12.5eV; it must have been emitted by the Ly$\gamma$ line at 12.75eV at
$z = 20.4$.  The fraction of photons that escape from these walls
during an integration step is
\begin{equation}
\label{lyman_abs}
f = \frac{1 - [(1+z)/(1+z_{screen})]^{1.5+\alpha}}
{1 - [(1+z)/(1+z+\Delta z)]^{1.5+\alpha}},
\end{equation}
where $\alpha$ = --1.8 is the slope of a power-law spectrum ($F_\nu
\propto \nu^{-\alpha}$) and
\begin{equation}
\label{abs_wall}
z_{screen} = \frac{\nu}{\nu_{Lyi}}(1+z_{obs}) - 1.
\end{equation}
$\nu_{Lyi}$ is the nearest, blueward Lyman transition to $\nu$.  Inherently,
$z_{screen}$ must be between $z$ and $z+\Delta z$.
\section{Results}
In this Section, we present the results of our calculation of the
evolution of SNe rates, volume-averaged metallicity, and UV
background.  We also present the variance of optical depth to electron
scattering and SNe rates with different protogalaxy star formation
scenarios.  In Figure \ref{evo_plot}, we plot minimum halo mass,
density of those halos, metallicity, SNe rate, and UV background in
the LW band for the main models.

\begin{figure}[!t]
\vspace{0.75cm}
\plotone{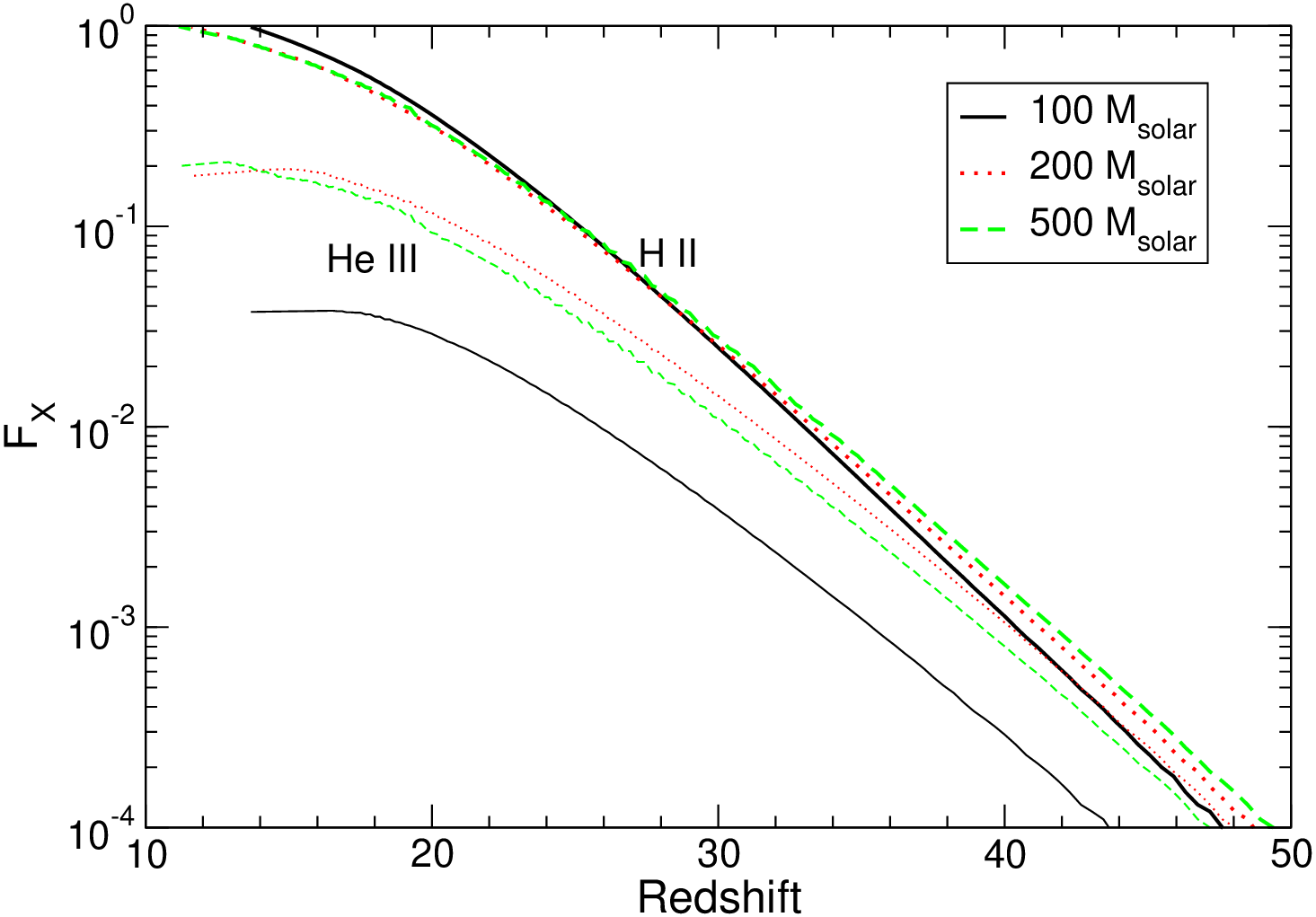}
\caption{\label{filling}
Evolution of filling factors of ionized hydrogen (top) and
doubly ionized helium (bottom).  The legend is the same as Figure
\ref{evo_plot}.}
\end{figure}
\subsection{Metallicity}
Metal-free stars with masses between $\sim$140--260$\msol$ result in a
SNe that is visible and eject heavy elements into the IGM.  When a
star is within this range, the stellar core has sufficient entropy
after helium burning to create positron/electron pairs.  These pairs
convert the gas energy into mass while not greatly contributing to
pressure.  This creates a major instability, where the star contracts
rapidly until oxygen and silicon implosive burning occurs.  Then the
star totally disrupts itself by these nuclear explosions
\citep{Barkat67, Bond84}.  Stars between $\sim$100--140$\msol$
experience this instability.  It is not violent enough to disrupt the
star, but pulsations and mass loss occur until equilibrium is reached
and the hydrogren envelope is ejected \citep{Bond84, Heger02a}.  Then
since it creates a large iron core, it probably forms a black hole,
preventing any further ejecta.  Additionally, metal-free stars above M
$\gsim$ 260$\msol$ may result in direct black hole formation, which
prohibits any significant heavy element ejecta and afterglow
\citep{Fryer01}.

For primordial stars in the pair-instability SNe mass range, we
determine a metallicity using our SNe densities and the metal
production of these SNe \citep{Heger02a}.  Except for iron,
significant spread in metallicity is not apparent for different
choices of primordial stellar mass.  Iron ejecta in pair instability
SNe range from 0.003--57$\msol$, so naturally we expect a large
spread.  For star formation in protogalaxies, we use carbon and metal
production values from \citet{Renzini81} for 1--8 $\msol$ and
\citet{Maeder92} for 9--120 $\msol$.  Additionally for 15--25$\msol$,
we use the iron nucleosynthesis yields from \citet{Rauscher02}.  At
low metallicities and above $\sim$40$\msol$, direct black hole
formation occurs after stellar collapse, which inhibits any ejecta
from escaping the remnant.  Furthermore above $\sim$25$\msol$, a black
hole forms through ``fallback'' of ejecta onto the neutron star
remnant \citep{Fryer99}.  Thus we set an upper mass cutoff of
25$\msol$ for metal enrichment from protogalaxies.  With the chosen
Salpeter IMF, 2.9 $\times$ 10$^{-3}$, 0.023 and 0.071 $\msol$ of iron,
carbon, and metals, respectively, are produced for every $\msol$ of
stellar material.  It should be noted that most of the carbon is
produced by intermediate mass stars, and its injection is delayed by a
few hundred million years, which corresponds to the first metal
enriched, intermediate mass stars exploding at z $\sim$ 12.
Furthermore, we ignore the delay time of any stellar lifetimes in
Figure \ref{metallicity} to give a rough estimate of the
volume-averaged metallicity evolution.

With the evolution of metallicity, we can estimate the period of
metal-poor (--5 $\lsim$ [Fe/H] $\lsim$ --1) star formation.  For the
most metal deficient stars, their metallicities of [Fe/H] = (--4.0,
--5.3) correspond to a formation epoch of z $\sim$ (28, 36)
\citep{Norris01, Christlieb02}.  Additionally, primordial stars in the
200$\msol$ model account for [Fe/H] $\sim$ --5.4 of the total iron
ejecta.  If we set the primordial stellar mass to be 250$\msol$, these
stars eject tremendous amounts of iron into the IGM and possibly
enhance [Fe/H] to --4.1.  Conversely, primordial stars at the lower
mass range of pair-instability SNe produce insignificant amounts of
iron and will not pollute the IGM.  The IGM iron contribution from
primordial stars is likely lower than these quoted values since in a
proper IMF not all stars will result in a pair instability SNe.

%
Possibly these metallicities correspond to the VMS (Very Massive Star)
component of metallicity of metal-poor (--4 $<$ Z/Z$_\odot$ $<$ --1)
stars \citep{Qian02}.  They assume that a VMS contributes
10$^{-4}$Z$_\odot$ to the IGM.  By inspection of the evolution of
metal yields from protogalaxies and primordial stars, the VMS
component may contain contributions from protogalaxies since
primordial stars are suppressed by the UV background before producing
a universal metallicity of 10$^{-4}$Z$_\odot$.  Lastly, metallicities
found in z $\sim$ 5 Lyman alpha clouds are [C/H] $\sim$ --3.7 and much
less than our calculated values \citep{Songaila01}.  The bulk of the
metals might not migrate to the void in which Ly$\alpha$ clouds exist,
and the host galaxies would retain the majority of the metals.  This
behavior may explain the lack of metals observed in Ly$\alpha$ clouds
and the higher than expected metallicity in the VMS component.
However, the clustering of the first objects, primordial IMF, and the
mixing of metals in the IGM will have to be taken into account before
we can predict metallicities of the Lyman alpha forest and metal
deficient stars with high confidence.

\begin{figure}[!t]
\vspace{0.75cm}
\plotone{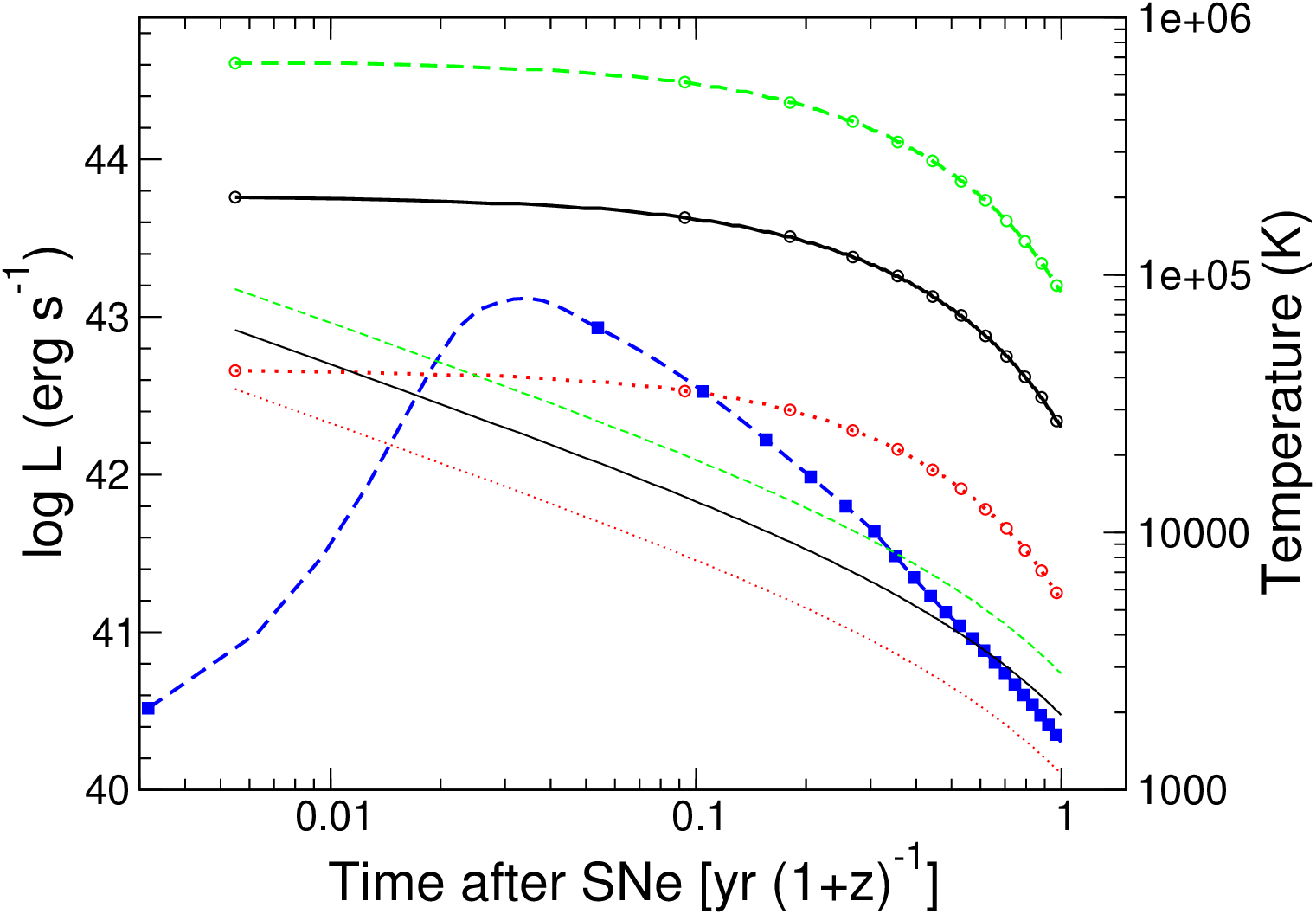}
\caption{\label{sne_lum}Using the decay of $^{56}$Ni and Equation
(\ref{sne_temp}), we calculate the luminosity of primordial SNe and
its effective temperature.  Heavy ({\em top}) lines with circles are
luminosities and light ({\em bottom}) lines are temperature.  The blue
dashed line with filled squares is a Type Ia luminosity evolution for
comparison \citep{Woosley86}.  The lines from top to bottom ({\em
dotted, solid, dashed}) are for stellar masses 175, 200, and 250
$\msol$.}
\end{figure}
\subsection{Optical Depth to Electron Scattering}
Optical depth due to electron scattering (\ref{thomson}) is a good
observational test to determine the neutral fraction of the Universe
before reionization.  The WMAP satellite measured $\tau_{es}$(z) =
0.17 $\pm$ 0.04 at 68\% confidence.
\begin{equation}
\label{thomson}
\tau_{es}(z) = \int^z_0 \bar{n}_e \sigma_{TH} c \left(\frac{dt}{dz}\right)
dz,
\end{equation}
where $n_e$ is the proper electron density and $\sigma_{TH}$ is the
Thomson cross-section.  To accurately calculate $n_e$, we must
consider all ionizations of primordial gas, which includes H$^+$,
He$^+$, and He$^{++}$.  Therefore, the proper electron density is
\begin{equation}
n_e = (1+z)^3 \left(n_{H}F_{H^+} + n_{He}F_{He^+} +
2n_{He}F_{He^{++}}\right),
\end{equation}
where F$_{H^+}$ F$_{He^+}$, and F$_{He^{++}}$ are the ionized volume
fraction of H$^+$, He$^+$, and He$^{++}$, respectively.  The effect of
more luminous primordial stars is evident in Figure \ref{filling} as
they ionize the Universe faster at high redshifts.  To match the WMAP
result, less ionizing photons are necessary from protogalaxies if the
primordial IMF is skewed toward higher masses.  Although these models
have the same total $\tau_{es}$, cosmological reionization occurs at z
= 13.7, 11.6, 11.2 for M$_{FS}$ = 100, 200, and 500$\msol$,
respectively.  However, these reionization redshifts are not
consistent with observed Gunn-Peterson troughs in z $\sim$ 6 quasars
\citep{Becker02, Fan02} and the high IGM temperatures at z $\sim$ 4
inferred from Ly$\alpha$ clouds \citep{Hui03}.  Perhaps portions of
the Universe recombine after complete reionization, which will match
the most distant quasar observations \citep{Cen02}.  If this were
true, the first reionization epoch has to be faster and earlier to
compensate for this partial recombination and to match the WMAP
result.  This would lower our SNe rates slightly since primordial star
formation will be further suppressed by the quicker reionization.

The ionizing history of the Universe is directly related to the number
of ionizing photons that are produced and escape into the IGM.  If we
fix $\tau_{es}$ = 0.17, we constrain f$_\star$ and f$_{esc}$ to a
power law
\begin{equation}
f_{esc} = B \: f_\star^{-a},
\end{equation}
where B = [0.00307, 0.00239, 0.00217] and a = [0.906, 0.839, 0.806]
for primordial stellar masses 100, 200, and 500$\msol$, respectively.
The flattening of the power law with increasing M$_{FS}$ indicates the
increasing ionizing contribution from primordial stars.

\begin{figure}[!t]
\begin{center}
\vspace{0.75cm}
\plotone{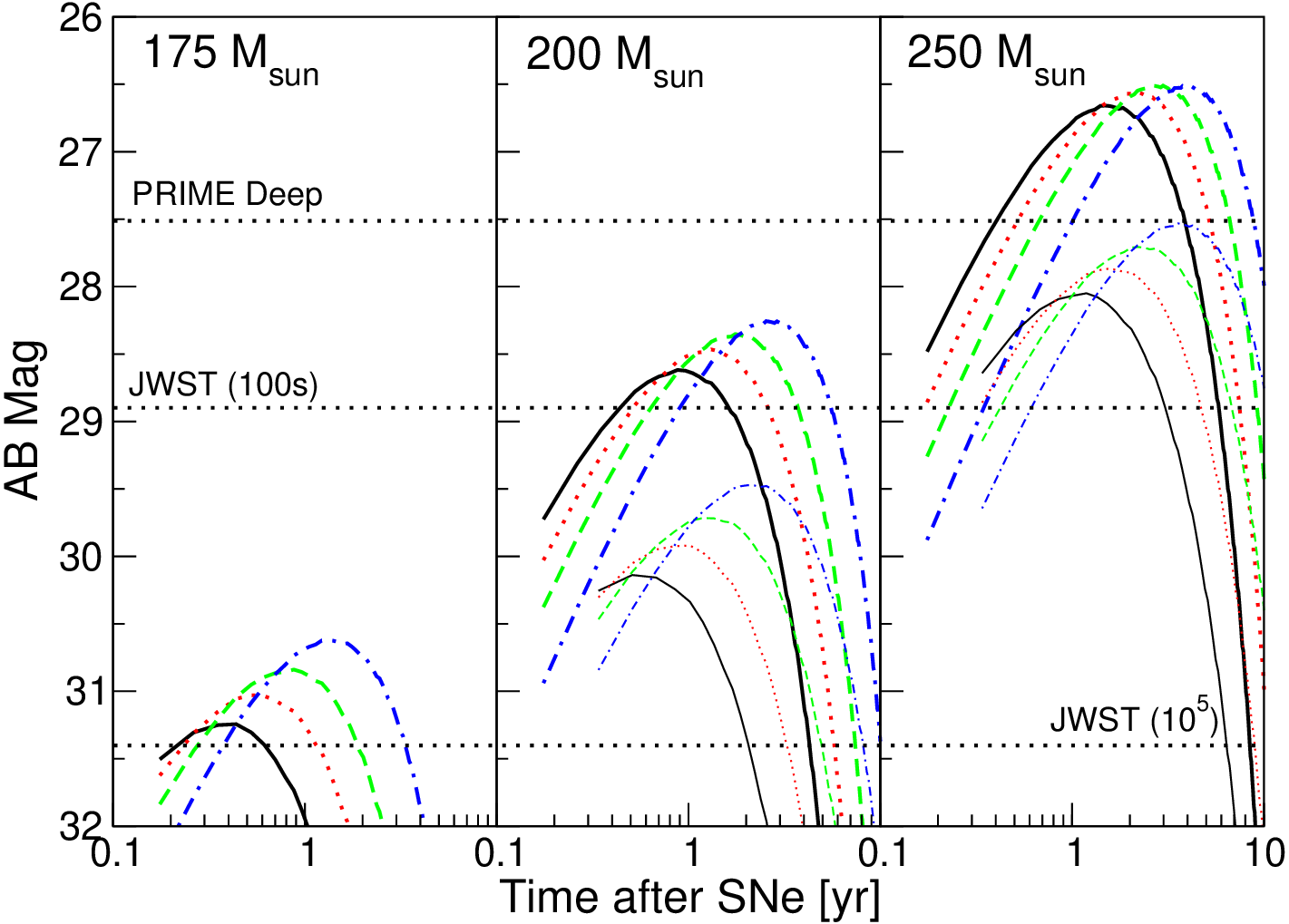}
\caption{\label{mags} ({\em Left to right}) Magnitudes for M$_{FS}$ =
175, 200, 250 $\msol$.  The dashed horizontal lines indicate limiting
magnitudes of space infrared observatories.  The heavy ({\em top}) and
light ({\em bottom}) lines are SNe at z = 15 and 30, respectively.
The magnitudes for the 175 $\msol$ and z = 15 case is not shown since
it is too dim.  The solid, dotted, dashed, and dash-dotted lines are
for Spitzer wavebands centered at (3.56, 4.51, 5.69, 7.96)
$\mu$m. {\em Note:} The detection limits of Spitzer and PRIME Medium
are not shown since they are too high at 24.5 and 25.6, respectively.}
\end{center}
\end{figure}
\begin{deluxetable}{ccccc}
\label{gaussFit}
\tablecolumns{5}
\tabletypesize{}
\tablewidth{0pc}
\tablecaption{Gaussian fits to M$_{{\rm min}}$ for the main models \\
  log(M) = A $\times$ exp[-(z-z$_0$)$^2$/(2$\sigma^2$)] + C}
\tablehead{\colhead{M$_\odot$} & \colhead{A} & \colhead{z$_0$} & \colhead{$\sigma$} & \colhead{C}}
\startdata
100 & 1.67 $\pm$ 0.01 & 10.0 $\pm$ 0.2 & 21.9 $\pm$ 0.1 & 5.19 $\pm$ 0.01 \\
250 & 1.71 $\pm$ 0.00 & 10.5 $\pm$ 0.0 & 23.3 $\pm$ 0.0 & 5.16 $\pm$ 0.00 \\
500 & 1.72 $\pm$ 0.01 & 11.3 $\pm$ 0.2 & 23.5 $\pm$ 0.1 & 5.15 $\pm$ 0.00
\enddata
\end{deluxetable}
\subsection{Primordial Supernovae Rates}
The natural units in our computation is comoving density, yet a more
useful unit is observed SNe yr$^{-1}$ deg$^{-2}$.  First we assume
these SNe are bright for 1 yr and then correct for time dilation.  We
consider the equation
\begin{equation}
  \label{rates}
  \frac{d^2N}{dtdz} = \frac{dV_c}{dz} \frac{d\rho_\star}{dt} (1 + z)^{-1},
\end{equation}
where the (1+z)$^{-1}$ converts the proper SNe rate into the observer
time frame, and
\begin{equation}
  \label{volume}
  \frac{dV_c}{dz} = D_H \frac{(1+z)^2 D_A^2}{E(z)} \: \Omega
\end{equation}
is the comoving volume element.  $\Omega$ (deg$^2$ = 3.046 $\times$
10$^{-4}$ sr) is the solid angle of sky that we want to sample.  $D_H
= c/H_0$ is the Hubble distance.  $D_A = D_M/(1+z)$ is the angular
diameter distance, and $D_M = D_H \int^z_0 E^{-1}(z') dz'$ is the
comoving distance \citep{Peebles93}.  The above equations are only
valid for a flat $\Lambda$CDM universe.

%
%
Since we only allow primordial stars to form in neutral regions, SNe
rates from these stars are highly dependent on the f$_{esc}$ and
f$_\star$, but less sensitive to the stellar primordial mass,
M$_{fs}$.  In our main models, the SNe rate varies little with
primordial stellar mass and is 0.34 yr$^{-1}$ deg$^{-2}$.  Even if we
vary f$_\star$ in a range 0.01--0.1 and fix $\tau_{es}$, primordial
SNe rates do not vary more than 10\% from the main models when
constrained by WMAP.  As f$_{esc}$ increases, the SNe rates decrease
due higher ionized volume filling factor.  The effect of a higher
f$_\star$ squelches SNe rates by two processes, a higher ionized
filling factor and higher UV background, which forces primordial stars
to form in more massive halos.

The primordial SNe rate peaks at z $\sim$ 12--20, later epochs for
larger primordial stellar masses, and falls sharply afterwards due to
the ensuing cosmological reionization.  Primordial star formation
ceases after z $\sim$ 12--16.  The combination of an increasing UV
background, reionizing Universe, and disruption of minihalos from
primordial SNe suppresses all primordial star formation.  For each
main model, we fit a Gaussian curve to the minimum mass of a minihalo
that can host a primordial star, and the parameters are listed in
Table 2 and are valid for z $>$ z$_0$.

It is fascinating that some rare primordial SNe occur at z $\gsim$ 40.
As an exercise, we estimate the earliest epoch of minihalo star
formation in the visible Universe with PS formalism and by considering
it takes $\sim$9.33 Myr for a halo to form a protostellar core
\citep{Abel02}.  With PS formalism, the ``first'' epoch equals where
the minihalo density is the inverse of the comoving volume inside z =
1000 (10523 Gpc$^3$).  Then we include an additional 9.33 Myr for the
ensuing star formation.  The halo masses of 1.74 $\times$ 10$^5$ and
10$^6$ $\msol$ correspond to formation times of z $\sim$ 71 and 64,
respectively.  Another interesting event to calculate is when the SNe
rate equals one per sky per year, which occurs at z $\sim$ 51 when
considering the collapse and star formation timescales.  This epoch is
in agreement with \citet{Escude03} estimate of z $\simeq$ 48.

\subsection{Magnitudes and Observability}
The magnitudes of primordial SNe are important as their occurrence
rates to catch these events unfolding in the distant universe.  We
exploit the $^{56}$Ni output from metal-free SNe models to calculate
luminosities L(t) from the two-step decay of $^{56}$Ni to $^{56}$Fe
\citep{Heger02a}.  We consider the emission spectrum to be a blackbody
spectrum with a time-dependent temperature (\ref{sne_temp}) calculated
using free expansion arguments.
\begin{equation}
\label{sne_temp}
T(t) = \left[\frac{L(t)M}{8\pi\sigma Et^2}\right]^{1/4} (1+z),
\end{equation}
where M is the stellar mass and E is the kinetic energy of the SNe.
The temperature and luminosity of a 175, 200, and 250$\msol$
primordial star SNe are depicted in Figure \ref{sne_lum}.  Kinetic
energy is taken from the SNe models \citep{Heger02a}.

In the upper range (M $>$ 200$\msol$) of pair-instability SNe, these
events produce 1--57$\msol$ of $^{56}$Ni \citep{Heger02a}, which will
produce tremendous luminosities compared to typical Type II SNe
outputs of only 0.1--0.4 $\msol$ \citep{Woosley86}.  However,
uncertainty in the primordial IMF places doubt in the frequency of
primordial star formation with high $^{56}$Ni yields.  Using
conventional Type II SNe parameters of L $\simeq$ 3 $\times$ 10$^{42}$
erg s$^{-1}$, T = 25000K (2 days $\lsim$ t $\lsim$ 7 days), and T =
7000K (7 days $\lsim$ t $\lsim$ 2 months) that are tuned by observed
light curves, \citet{Jordi97} determine apparent magnitudes that are
1--3 mag lower than our values, which is due to the greater nickel
production of new pair-instability SNe models.  Their model should
apply to the pair-instability SNe with little $^{56}$Ni ejecta.  The
temperatures are higher than our values because we allow for free
expansion of the ejecta due to a lower external medium gas density.
Radiation from the primordial star should expel most of surrounding
medium to create a low density, highly ionized region of approximately
100 pc in size for a 120$\msol$ and 500$\msol$ primordial star,
respectively \citep{Whalen04}.

In typical Type II light curves, radioactive decay does not
significantly contribute to the luminosity in the plateau stage
\citep{Popov93}.  In primordial SNe, the $^{56}$Ni decay may overwhelm
the typical sources of energy within the expanding fireball and create
a totally different light curve.  We stress that our light curve is a
very rough estimate at the processes occurring within a primordial
SNe.

In Figure \ref{mags}, we compare maximum AB magnitudes of low- and
high-mass pair-instability SNe at various redshifts and sensitivities
of space infrared observatories.  We consider the sensitivities of
SIRTF at 3.5\micron, PRIME medium and deep surveys, and JWST with
exposure times of 100s and 10$^5$ s.  We also give the typical light
curve for a Supernova Type Ia. Clearly our simple estimate leads to
very high predicted luminosities and more detailed numerical models of
these explosions are clearly desirable.

\begin{figure}[!t]
\begin{center}
\vspace{0.75cm}
\plotone{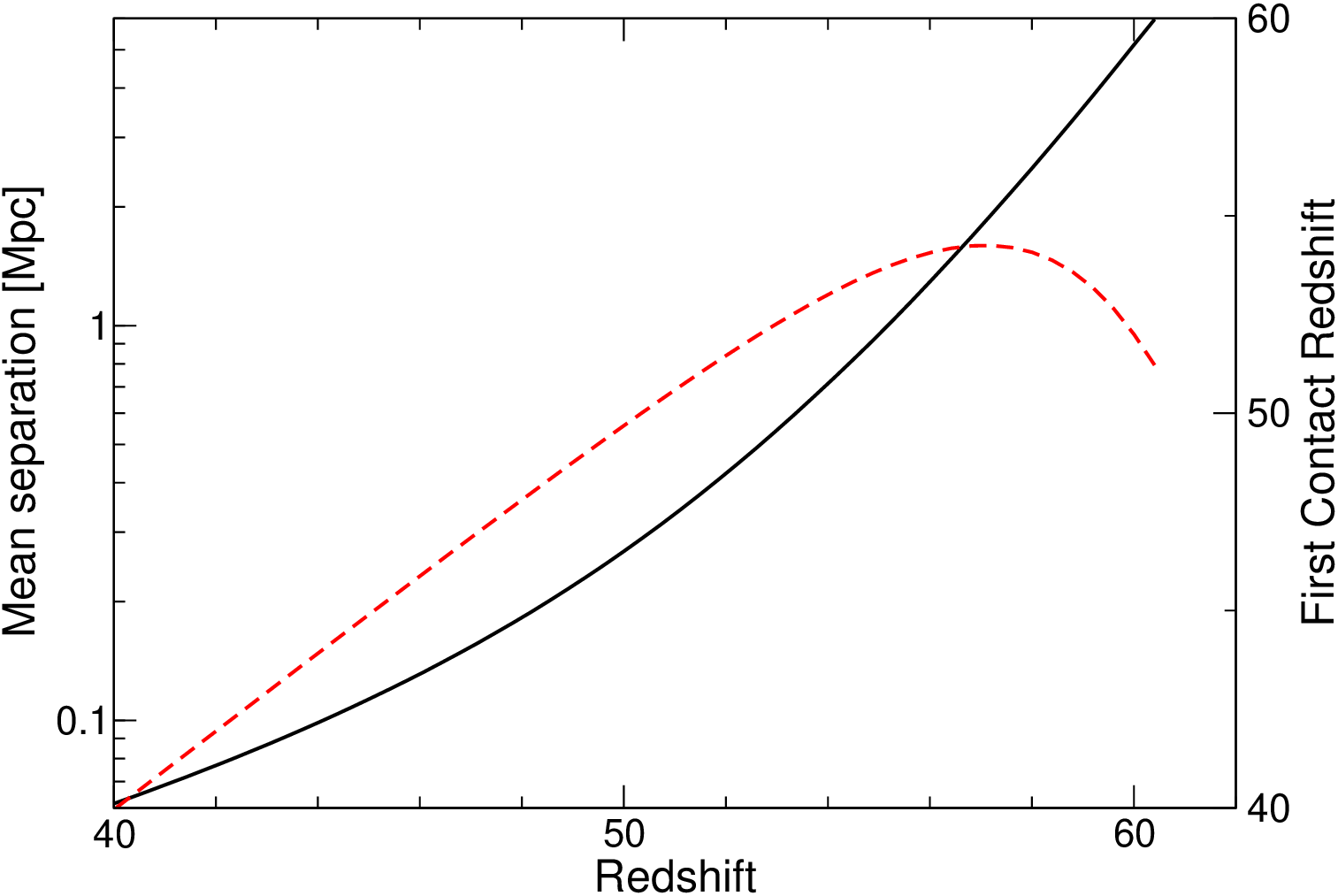}
\caption{\label{contact} The {\em solid} line is the mean proper
  separation between minihalos.  The {\em dashed} line is the redshift
  where minihalos first receive radiation from neighboring sources
  forming at the same epoch.}
\end{center}
\end{figure}
\subsection{Other factors}
Other factors could alter the feasibility of observing primordial
SNe. For instance, $\sim$2\% of primordial stars might die in
collapsar gamma-ray bursts (GRBs) \citep{Heger02b}.  \citet{Lamb99}
also provides a ratio of GRBs to Type Ib/c SNe rates that is $\sim$3
$\times$ 10$^{-5}$ ($f_{beam}$ / 10$^{-2}$)$^{-1}$.  However, this
estimate is for a normal Salpeter IMF, and as mentioned before, the
primordial star IMF could be skewed toward high masses, which would
increase the probability of a GRB.  Zero- and low-metallicity, massive
stars outside the pair-instability mass range die as GRBs or
jet-driven SNe, which are similar to GRBs but not as energetic and
spectrally hard.  If we consider a proportionality constant and
beaming factor, we estimate that the all-sky primordial GRB rate,
\begin{equation}
R_{GRB} = 2.8 \left(\frac{\theta_b}{0.01}\right)
\left(\frac{f_{GRB}}{0.02}\right) \; {\rm GRBs} \; {\rm yr}^{-1}.
\end{equation}
In M $\lsim$ 140$\msol$ stars, a collapsar results when it forms a
proto-neutron stars, cannot launch a supernova shock, and directly
collapses to form a black hole in $\sim$1 s \citep{Woosley93,
MacFadyen99}.  In M $\gsim$ 260$\msol$ stars, it does not create a
proto-neutron star and directly collapses into a massive black hole
\citep{Fryer01}.  Afterwards, these black holes can accrete gas and
produce X-rays that can further ionize the Universe
\citep{Ricotti03b}.  A few of these X-ray sources could be detected in
the Chandra deep field \citep{Ricotti04, Alexander03}.  In most GRB
models \citep[for an overview, see][]{Meszaros02}, the radiation is
beamed to small opening angles due to relativistic effects, which
would render a fraction of GRBs to be unobservable in our perspective.
Finally, gravitational lensing will significantly increase the
apparent magnitudes for selected primordial SNe for very small survey
areas.  However, the overall magnitude distribution is slightly dimmed
by gravitational lensing \citep{Marri98, Marri00}.

Using the proper minihalo density, we calculate the average light
travel time between sources, which indicates when negative \h2
feedback first affects star formation.  Star formation occurs
$\sim$9.33 Myr after halo formation, so radiation escapes into the IGM
at a time t$_{\rm{rad}}$ = t$_{\rm{H}}$(z$_{\rm{form}}$) + 9.33 Myr,
where z$_{\rm {form}}$ is the halo formation redshift.  These sources
have a mean proper separation of d = $\rho_\star^{-1/3}$ / (1+z);
therefore the time where radiation influences other halos is
t$_{\rm{rad}}$ + d/c.  The minimum of this time, corresponding to z =
54, is where radiation reaches other halos for the first time.  The
mean proper separation and epoch of first radiation effects are
illustrated in Figure \ref{contact}.

Deviations in the amplitudes of fluctuations, $\sigma_8$, and a
running spectral model \citep{Liddle92, Liddle93} also affect
primodial SNe rates.  We ran a set of models with $\sigma_8$ = 0.8,
and predictably, the rates decreased by a factor of $\sim$2.5 due to
lesser powers at small mass scales.  When we match $\tau$ = 0.17,
rates are 0.12--0.2 yr$^{-1}$ deg$^{-2}$ with the rates peaking
earlier at z $\sim$ 20, but now the redshift peaks are nearly
independent of redshift with only $\delta$z $\approx$ 1 separating the
100 and 500$\msol$ models.  If we keep the protogalaxy parameters from
the main models, rates do not change; however, the rates peak later
than our main models by $\delta$z $\approx$ 4, and $\tau$ lowers to
0.14.  According to inflationary models, the spectral index of
flucutations should be slowly varying with scale.  While analyzing
WMAP data, \citet{Peiris03} determined that the flucutation amplitude
is significantly lower at small scales.  With lesser powers at small
scales, primordial SNe rates decrease such as the case of a lower
$\sigma_8$.  In principle, studying primordial SNe occurrences with
respect to redshift could furnish direct constraints on the power
spectrum at these small scales.

\section{Constraining the Free Parameters}
We have demonstrated our model's dependence on the free parameters,
$f_{esc}$, $f_\star$, and M$_{fs}$.  We suggest several observational
and numerical methods in order to constrain our models.

\subsection{Observational}
To constrain early star formation history of dwarf irregulars with the
[$\alpha$/Fe] {\em vs.} [Fe/H] comparison, metal analyses of stars
with [Fe/H] $<$ --1.5 are needed to determine early star formation
rates of particular systems \citep[See Figure 1.5 in][]{Venn03}.
Dwarf spheroidals exhibit greater variance in star formation histories
from system to system, but a similar study will increase our knowledge
of high redshift star formation in these small galaxies to further
constrain $f_\star$ at high redshifts.  To improve on these studies,
the advances in multi-object spectrographs (e.g.
VIMOS\footnote{http://www.eso.org/instruments/vimos/},
FLAMES\footnote{http://www.eso.org/instruments/flames/}, and
GMOS\footnote{http://www.gemini.edu/sciops/instruments/gmos/gmosIndex.html})
enable the gathering of many stellar spectra in one exposure.  This
will greatly increase the stellar population data of dwarf galaxies.

Since some massive stars die as a long-duration GRB, these events can
convey information from the death of both Pop II and III stars.  The
propagation of the initial burst and afterglow provide information
about the total energy, gas density in the vicinity, and the Lorentz
factor of the beam.  The host galaxy ISM properties will help
constrain the $f_{esc}$ in high redshift galaxies.  As in the case of
GRB 030329 \citep{Stanek03}, the power-law spectra of the afterglow
can be subtracted to obtain a residual that resembles a typical SN
spectrum, which may be used to roughly determine the mass of the
progenitor.  Observing the afterglow is necessary to determine its
redshift.  The prospect of observing prompt afterglows will be
accomplished easier with {\em
Swift}\footnote{http://swift.gsfc.nasa.gov/}, which can possibly
detect GRB afterglows to z = 16 and 33 in the K and M bands,
respectively \citep{Gou04}.  Furthermore, $\gsim$50\% of GRBs occur
earlier than z = 5, and 15\% of those high redshift GRBs are
detectable by {\em Swift} \citep{Bromm02b}.  The comparison to nearby
SNe can provide crucial information of the high redshift ISM.
Finally, it is a possiblity to explore the intervening absorption with
the fast-pointing and multi-wavelength observations of {\em Swift}
\citep{Vreeswijk03, Loeb03, Barkana04}.  This IGM absorption would
constrain the reionization history of the Universe better, which may
change the primordial SNe rates, but more specifically the rate per
unit redshift which would roughly conform to the filling factor
evolution.

The radiation from protogalaxies and primordial stars will not only
ionize the Universe but also contribute to the near-infrared
background (NIRB).  Calculations have shown that primordial stars can
provide a significant fraction of radiation to the NIRB
\citep{Santos02, Salvaterra03}; however, the paradigm of early
reionization set by WMAP was not considered at the time.  f$_\star$ of
protogalaxies and densities of primordial stars can be further
constrained if future studies of the NIRB consider the large
emissivities of zero- and low-metallicity sources at z $\gsim$ 15.

Other outlooks include detecting high redshift radio sources and
searching for 21cm absorption and emission \citep{Hogan79, Scott90,
Iliev03, Furlanetto04}.  By looking for 21cm signatures, observations
would be directly probing the neutral regions of the Universe since
the Gunn-Peterson trough saturates only at a neutral filling factor of
$\sim$10$^{-5}$.  Lastly, additional searches for high redshift
starbursts \citep[e.g.][]{Ellis01, Pello04} in lensed fields will
furnish an understanding of the characteristics of these objects and
help tighten our models of early ``normal'' and primordial star
formation.

\subsection{Numerical Simulations}
Metal production from primordial stars cannot account for the volume
filling factor of metals as seen in Ly$\alpha$ clouds in current
models \citep{Norman04}.  Therefore, ubiquitous star formation in
protogalaxies most likely polluted the sparse regions of the Universe.
Combining the primordial and protogalaxy metal output, simulations
with proper metal transport should agree with the abundances observed
in Ly$\alpha$ clouds.  Such simulations that takes into account both
of these metal sources and matches the results of \citet{Songaila96}
and \citet{Songaila01} is needed to constrain star formation before
reionization.

Numerical simulations are also needed to investigate radiative
transfer from a protogalaxy stellar population.  This scenario will
contain more complexities than a single primordial star in a spherical
halo as in \citet{Whalen04}.  However a protogalactic simulation with
the Jeans length resolved, star formation and feedback, and radiative
transfer, it will be possible to study the evolution of the ISM in a
protogalaxy, which will produce insight and better constraints on the
photon escape fraction, $f_{esc}$, in the Lyman continuum at high
redshifts.  \citet{Ricotti02a, Ricotti02b} thoroughly study radiative
transfer around protogalaxies; however, $f_{esc}$ is still a parameter
when it should be determined from the radiative transfer results in
the simulations.  Ideally, the analytical ideas about ionizing the
Universe in \citet{Madau95, Haardt96, Madau99} should be realized in
such simulations but in the context of the current paradigm of a high
redshift reionization as indicated by WMAP.  Also we may hope by using
numerical radiative transfer techniques for line and continuum
radiation in three dimension to push the simulations of \citet{Abel02}
to follow the entire accretion phase of the first stars.  This would
lead to stronger constraints on the masses of the very first stars.

\section{Summary and Conclusions}
The WMAP measurement of optical depth to electron scattering places a
constrain of early cosmological reionization of the universe, which we
show to be mainly from star formation in protogalaxies.  This general
result is in agreement with other studies of reionization after WMAP
\citep[e.g.][]{Cen03, Somerville03, Ciardi03, Ricotti03a, Sokasian04}.
\begin{itemize}
\item The radiation from protogalaxies squelches out primordial star
formation, and {\em $\sim$0.34 primordial SNe deg$^{-2}$ yr$^{-1}$} are
expected.  SNe rates can vary from 0.1 to $>$1.5 deg$^{-2}$ yr$^{-1}$
depending on the choice of primordial stellar mass and protogalaxy
parameters while still constrained by the WMAP result.  The peak of
SNe rate occurs {\em later} with {\em increasing primordial stellar
masses}.  These results are upper limits since the rate of visible
primordial SNe depends on the IMF because only a fraction will lie in
the pair instability SNe mass range.  The other massive primordial
stars might result in jet-driven SNe or long duration GRBs.
\item Stellar abundances and star formation in local dwarf galaxies
aid in estimating protogalaxy characteristics.  We choose f$_\star$ =
0.04 and f$_{esc}$ = 0.050 in our 100 $\msol$ model.  In protogalaxies,
the star formation efficiency is slightly lower than local values, but
the photon escape fraction is within local observed fractions.
\item Primordial stars enrich the IGM to a maximum volume-averaged
[Fe/H] $\simeq$ --4.1 if the IMF is skewed toward the pair-instability
upper mass range.  A proper IMF will lessen this volume-averaged
metallicity since only a fraction of stars will exist in this mass
range.  If complete metal mixing does not occur, there will be
metal-rich and metal-poor regions relative to the universal
metallicity.
\item The entire error bar of the WMAP measurement of optical depth to
electrons can be explained by a higher/lower f$_\star$ and f$_{esc}$.
Protogalaxies can ionize the IGM easily since low metallicity
starburst models produce 20--80\% more ionizing photons as previously
used (Z = 0.001) IMFs.  Massive primordial stars provide $\sim$10\% of
the necessary photons to achieve reionization.  No exotic processes or
objects are necessary.
\item Only the upper mass range of pair instability SNe will be
observable with JWST since the low mass counterparts do not produce
enough $^{56}$Ni to be very luminous.
\end{itemize}

Although the IMF of primordial stars is unknown, simulations have
hinted that a fraction of metal-free stars may exist in the
pair-instability mass range.  When observational rates, light curves,
and spectra are obtained from future surveys, these data would provide
very stringent constraints on the underlying CDM theory as well as our
understanding of primordial star formation.

\acknowledgments{This work was supported by NSF CAREER award
AST-0239709 from the National Science Foundation.  JHW also
acknowledges support from Pennsylvania Space Grant.  We are grateful
to Alexander Heger for useful discussions about primordial SNe
luminosities.  We are also grateful to Leonidas A. Moustakas for
encouraging us to investigate the feasibility of a future detection.}

\end{document}